\documentclass{bmcart}

\usepackage{amsmath}
\usepackage{graphicx}
\usepackage{hyperref}
\usepackage[ruled]{algorithm2e}
\definecolor{goodblue}{RGB}{0, 91, 187}
\hypersetup{
  colorlinks=true,
  allcolors=goodblue,
  urlcolor=goodblue,
  citecolor=goodblue,
  pdfborder={0 0 0},
  breaklinks=true,
}
\usepackage[utf8]{inputenc} %unicode support

\startlocaldefs

\newcommand{\bH}{\mathbf{H}}
\newcommand{\bS}{\mathbf{S}}
\newcommand{\ssf}{\sigma_{\mathrm{SF}}}
\newcommand{\ses}{\sigma_{\mathrm{ES}}}
\newcommand{\sfes}{\sigma_{\mathrm{FES}}}
\newcommand{\des}{d_{\mathrm{ES}}}
\newcommand{\dnes}{d_{\mathrm{NES}}}

\newcommand{\ER}{Erd\"os-R\'enyi }
\newcommand{\ps}[1]{\mathcal{P}(#1)}
\newcommand{\srps}[2]{\mathcal{P}_{#2}(#1)}

\newcommand*{\Break}{\textbf{break}}

\newlength\customlen
\newcommand\custominput[1]{%
  \settowidth\customlen{\KwIn{}}%
  \setlength\hangindent{\customlen}%
  \hspace*{\customlen}#1\\}
\endlocaldefs

\begin{document}

%%% Start of article front matter
\begin{frontmatter}

\begin{fmbox}
\dochead{Research}


\title{The simpliciality of higher-order networks}

%%%%%%%%%%%%%%%%%%%%%%%%%%%%%%%%%%%%%%%%%%%%%%
%%                                          %%
%% Enter the authors here                   %%
%%                                          %%
%% Specify information, if available,       %%
%% in the form:                             %%
%%   <key>={<id1>,<id2>}                    %%
%%   <key>=                                 %%
%% Comment or delete the keys which are     %%
%% not used. Repeat \author command as much %%
%% as required.                             %%
%%                                          %%
%%%%%%%%%%%%%%%%%%%%%%%%%%%%%%%%%%%%%%%%%%%%%%

\author[
   addressref={aff1,aff2},                   % id's of addresses, e.g. {aff1,aff2}
   corref={aff1},                       % id of corresponding address, if any
   email={nicholas.landry@uvm.edu}   % email address
]{\inits{NWL}\fnm{Nicholas W.} \snm{Landry}}
\author[
   addressref={aff1,aff2},
   email={jean-gabriel.young@uvm.edu}
]{\inits{JGY}\fnm{Jean-Gabriel} \snm{Young}}
\author[
   addressref={aff3},
   email={eikmeier@grinnell.edu}
]{\inits{NE}\fnm{Nicole} \snm{Eikmeier}}

%%%%%%%%%%%%%%%%%%%%%%%%%%%%%%%%%%%%%%%%%%%%%%
%%                                          %%
%% Enter the authors' addresses here        %%
%%                                          %%
%% Repeat \address commands as much as      %%
%% required.                                %%
%%                                          %%
%%%%%%%%%%%%%%%%%%%%%%%%%%%%%%%%%%%%%%%%%%%%%%

\address[id=aff1]{%                           % unique id
  \orgname{Vermont Complex Systems Center, University of Vermont}, % university, etc
  \street{82 Innovation Pl},                     %
  \postcode{05405}                                % post or zip code
  \city{Burlington},                              % city
  \cny{USA}                                    % country
}
\address[id=aff2]{%
  \orgname{Department of Mathematics and Statistics, University of Vermont},
  \street{82 Innovation Pl},
  \postcode{05405}
  \city{Burlington},
  \cny{USA}
}
\address[id=aff3]{%
  \orgname{Department of Computer Science, Grinnell College},
  \street{1116 8th Ave},
  \postcode{50112}
  \city{Grinnell},
  \cny{USA}
}

\end{fmbox}

%%%%%%%%%%%%%%%%%%%%%%%%%%%%%%%%%%%%%%%%%%%%%%
%%                                          %%
%% The Abstract begins here                 %%
%%                                          %%
%% Please refer to the Instructions for     %%
%% authors on http://www.biomedcentral.com  %%
%% and include the section headings         %%
%% accordingly for your article type.       %%
%%                                          %%
%%%%%%%%%%%%%%%%%%%%%%%%%%%%%%%%%%%%%%%%%%%%%%

\begin{abstractbox}

\begin{abstract} % abstract

Higher-order networks are widely used to describe complex systems in which interactions can involve more than two entities at once.
In this paper, we focus on inclusion within higher-order networks, referring to situations where specific entities participate in an interaction, and subsets of those entities also interact with each other.
Traditional modeling approaches to higher-order networks tend to either not consider inclusion at all (e.g., hypergraph models) or explicitly assume perfect and complete inclusion (e.g., simplicial complex models).
To allow for a more nuanced assessment of inclusion in higher-order networks, we introduce the concept of "simpliciality" and several corresponding measures.
Contrary to current modeling practice, we show that empirically observed systems rarely lie at either end of the simpliciality spectrum.
In addition, we show that generative models fitted to these datasets struggle to capture their inclusion structure.
These findings suggest new modeling directions for the field of higher-order network science.

\end{abstract}

%%%%%%%%%%%%%%%%%%%%%%%%%%%%%%%%%%%%%%%%%%%%%%
%%                                          %%
%% The keywords begin here                  %%
%%                                          %%
%% Put each keyword in separate \kwd{}.     %%
%%                                          %%
%%%%%%%%%%%%%%%%%%%%%%%%%%%%%%%%%%%%%%%%%%%%%%


\end{abstractbox}

\end{frontmatter}

%%%%%%%%%%%%%%%%%%%%%%%%% start of article main body

\section{Introduction}
\label{sec:introduction} 

A wide range of complex systems are shaped by interactions involving several entities at once: social networks are driven by group behavior \cite{mastrandrea_contact_2015}, emails often have multiple recipients \cite{benson_simplicial_2018,klimt_enron_2004,leskovec_graph_2007}, molecular pathways in cells involve multi-protein interactions \cite{murgas_hypergraph_2022}, and scientific articles involve groups of co-authors \cite{patania_shape_2017}. 
Higher-order networks are a natural extension to networks explicitly designed to model such multi-way relationships~\cite{battiston_networks_2020}.

Two mathematical representations are most commonly used to model higher-order networks: hypergraphs and simplicial complexes~\cite{torres_why_2021}.
A hypergraph is a collection of entities (nodes) connected by interactions (hyperedges) between any number of these entities.
A simplicial complex can be considered a hypergraph with an additional requirement known as \emph{downward closure}, which states that when an interaction exists between $m$ entities, every possible sub-interaction also exists.
This mathematical construction originates in algebraic topology and is motivated by theoretical applications; for example, forming operators such as boundary matrices to identify cycles in a dataset or the Hodge Laplacian to describe dynamical processes in higher-order networks \cite{eckmann_harmonische_1944,bianconi_higher-order_2021}.

Recent work has grappled with the problem and consequences of choosing the proper representation---simplicial complex, hypergraph, or other---for a given complex system. 
Ref.~\cite{zhang_higher-order_2023}, for instance, shows that synchronization can differ drastically in systems modeled with simplicial complexes and hypergraphs due to synchrony driven by the included edges of simplicial complexes, and three recent studies investigate the impact of inclusions on contagion~\cite{kim_contagion_2023,larock_encapsulation_2023,burgio_triadic_2024}.
Additionally, Ref.~\cite{torres_why_2021} discusses how each representation corresponds to different modeling assumptions and, thus, different analysis pipelines.

Missing from these studies are analyses of the suitability of higher-order representations for describing empirically observed interactions.
When a set of interactions is given to a data scientist or modeler, the choice of representation is essentially empirical.
Do the data satisfy downward closure?
(In which case, a simplicial complex may best represent it.) 
Or do the data violate downward closure?
(In which case, it should be modeled as a hypergraph.)

In this paper, we introduce the concept of \emph{simpliciality} to describe the extent to which a set of interactions satisfies the downward closure requirement.
We implement this concept with three measures of the overall simpliciality of a dataset and describe how to define local versions of these global metrics.
Using these measures, we investigate the simpliciality of empirical datasets and show that commonly analyzed higher-order interaction datasets populate the full spectrum of simpliciality.
We find that there may be large variations in the simpliciality depending on the chosen empirical dataset and the measure of simpliciality.
Additionally, we show that the level of simpliciality displayed by existing models is typically not captured by existing generative models for higher-order networks.
Hence, this paper identifies an essential gap in the current set of higher-order measures and models.

These new simpliciality measures complement other higher-order structural measures such as community structure \cite{chodrow_generative_2021,zhou_learning_2006,kaminski_clustering_2019}, centrality \cite{benson_three_2019,feng_hypergraph_2021,tudisco_node_2021}, clustering \cite{gallagher_clustering_2013,klimm_hypergraphs_2021}, assortativity \cite{chodrow_configuration_2020,landry_hypergraph_2022}, and degree heterogeneity \cite{landry_effect_2020}. 
While these measures are helpful in understanding how higher-order data is organized, they do not address how hypergraphs relate to simplicial complexes. 
Closest to our work is the recent Ref.~\cite{larock_encapsulation_2023}, in which the authors describe the encapsulation graph, a structural description of the inclusion patterns of any given dataset, as well as a dynamical process based on inclusion that spreads from included hyperedges to larger containing hyperedges.
Also relevant is Ref.~\cite{joslyn_hypernetwork_2021}, which defines a global metric of inclusions.
There is a large body of literature surrounding the concept of \textit{nestedness}~\cite{mariani_nestedness_2019}, which measures the inclusion structure of unipartite and bipartite networks, particularly in ecological contexts~\cite{bastolla_architecture_2009}.
Measures of nestedness, however, do not use simplicial complexes as a reference point against which to compare.
In contrast, our approach describes downward inclusions succinctly with simple global measures,  offering a new perspective on how higher-order data is organized.

% ==========================================
\subsection{Mathematical definitions}
\label{subsec:definitions} 
% ==========================================

We encode interactions as \emph{hypergraphs}, defined as a pair $\bH = (V, E)$ where $V$ is a set of $N=|V|$ entities known as nodes, and where $E$ is a set of subsets of $V$ encoding relationship between nodes.
We refer to a set $e\in E$ as a "hyperedge" or just "edge," and define the \emph{size} of an edge as $|e|$.
In general, $E$ could be a multiset of multisets, but in this study, we solely consider simple hypergraphs, where each edge is only present once (no multi-hyperedges), and each edge can only contain unique entities (no self-relations).

Our analysis will focus on the prevalence of \emph{inclusions} in interaction data, so we describe this relationship formally.
We say that an edge $f$ is included in $e$, or $f \subset e$, if every node of $f$ is also a node of $e$. 
Drawing from the nomenclature of algebraic topology, we also say that $f$ is a \emph{subface} of $e$~\cite{hatcher_algebraic_2001}.

Inclusions naturally lead to the concept of a \emph{maximal hyperedge}, i.e., a hyperedge $\widetilde{e}$ that is not included in any other hyperedge.
We denote the set of all the maximal hyperedges of $\bH$ as $\widetilde{E}(\bH)$. 

A \emph{simplex} $s$ is then a collection of hyperedges that contains a single maximal hyperedge $\widetilde{e}$ and satisfies $s \equiv \ps{\widetilde{e}}$, where $\ps{X}$ is the power set of set $X$.
In other words, a simplex is a maximal edge with an associated collection of subfaces in which every possible subface of the maximum edge exists. 
A collection of simplices is a \emph{simplicial complex} $\bS=(V, E)$, and we say that a hypergraph where every maximal edge is a simplex satisfies \emph{downward closure}.

Finally, we will use the notion of an \emph{induced simplicial complex}, which is the simplicial complex constructed from the maximal edges of a hypergraph by adding all hyperedges needed to satisfy downward closure.

% ==========================================
\section{Measuring simpliciality}
% ==========================================

\begin{figure}
    \centering
    \includegraphics[width=8.6cm]{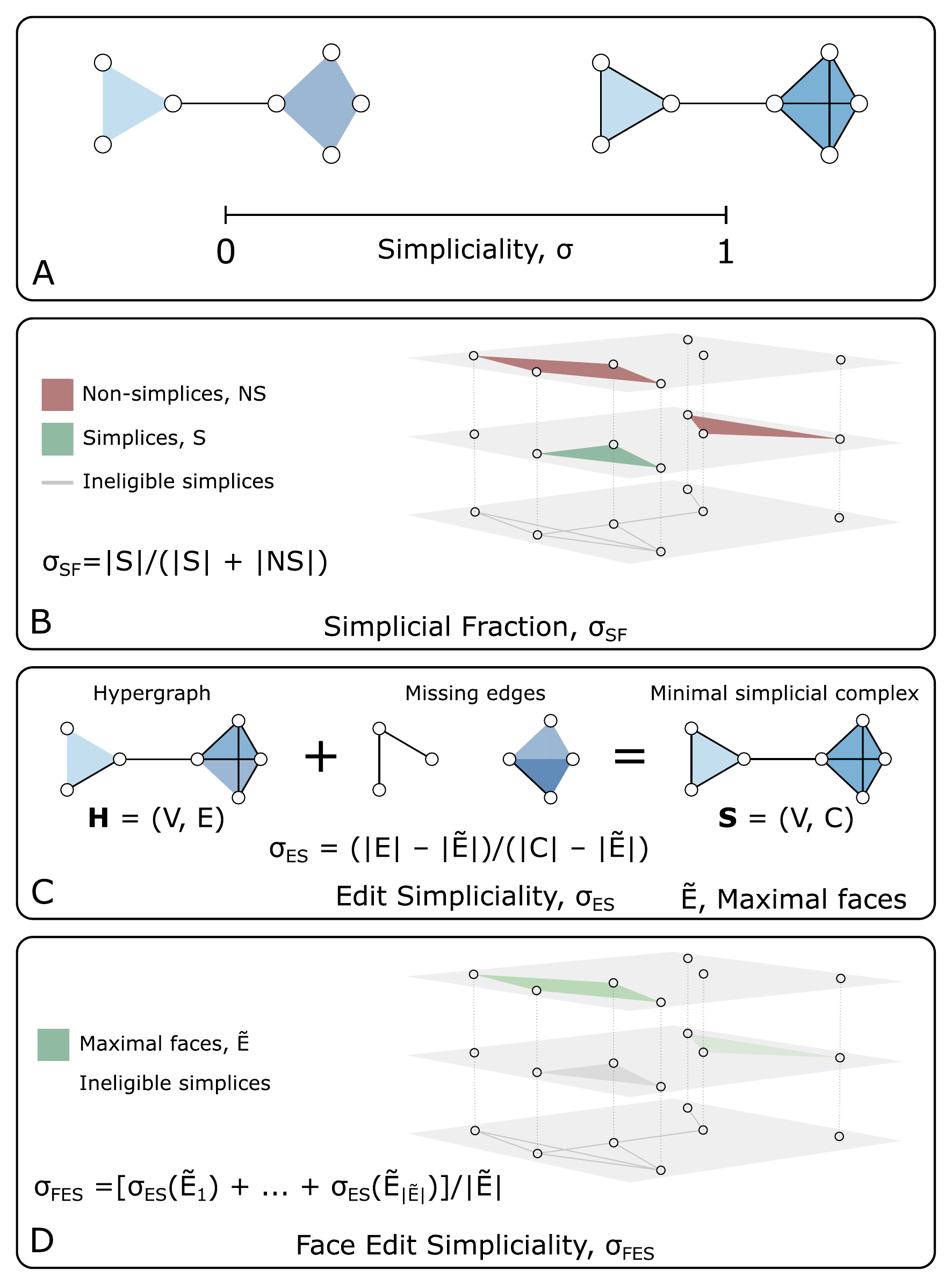}
    \caption{\csentence{An illustration of simpliciality.} Simpliciality captures the notion of inclusion in higher-order networks. 
    (A) Interaction data are fully simplicial, denoted $\sigma=1$, when all possible included interactions exist, e.g., an interaction between three nodes appears together with interactions between all three possible pairings of these nodes. 
    Data are minimally simplicial, denoted $\sigma=0$, when no included interactions are present.
    (B-D) Three natural measures of simpliciality which place higher-order datasets on the simplicial spectrum between $\sigma=0$ and $\sigma=1$, as described in Sec.~\ref{subsec:measures}.}
    \label{fig:explanation_figure}
\end{figure}

This paper introduces the concept of \emph{simpliciality}. 
Simpliciality, broadly defined, measures the inclusion structure of a hypergraph and how similar a higher-order dataset is to the structure of a simplicial complex; see Fig.~\ref{fig:explanation_figure}A. 
There are many ways to measure this, which we outline in Sec.~\ref{subsec:measures}.
Before we get there, however, we must first introduce relevant terminology.

% ==========================================
\subsection{Measures} 
\label{subsec:measures}
% ==========================================

This section introduces measures of simpliciality.
We follow a few guiding principles to design these measures.
One, a simplicial complex must be maximally simplicial with respect to any measure of simpliciality. 
Two, to facilitate easier comparison between datasets, measures of simpliciality should be normalized so that they map a hypergraph to a value between 0 and 1. 
Three, if a subface is added to a hypergraph, the simpliciality must increase. 
Four, as a dataset becomes qualitatively more like a simplicial complex, the simpliciality should increase. 
And five, we stipulate that the simpliciality of an empty hypergraph is undefined.

There are many ways to define a measure of simpliciality while maintaining these guiding principles. To highlight different structural elements contributing to the inclusion structure, we define three measures: the simplicial fraction, the edit simpliciality, and the mean face simpliciality. These measures are all illustrated in Fig.~\ref{fig:explanation_figure}~B-D.

\vspace{0.1in}
\noindent \textbf{Simplicial Fraction}.
In a simplicial complex, every subface is itself a simplex, so when a hypergraph is a simplicial complex, it contains all subsets of each of its hyperedges. 
The \emph{simplicial fraction} (SF) measures the degree to which this is true, defined as the fraction of hyperedges which are also simplices. 

Formally, we let $\bH = (V, E)$ be a hypergraph and let $S = \{ e\in E \mid \ps{e}\subseteq E\}$ be the set of hyperedges which are also simplices. Then, the simplicial fraction is defined as 
\begin{equation}
\ssf = \frac{|S|}{|E|}
\end{equation}
and it takes values in the range $\ssf\in[0,1]$; see Fig.~\ref{fig:explanation_figure}B.

The simplicial fraction directly measures the number of simplices in the dataset and is, therefore, highly interpretable.
However, one potential downside is that edges which \emph{almost} achieve downward closure do not count at all toward the overall simpliciality.
Furthermore, this definition weighs smaller simplices heavily, as small simplices contribute to the simpliciality of all hyperedges that include them.

\vspace{0.1in}
\noindent \textbf{Edit Simpliciality}.
The \emph{edit simpliciality} (ES) is defined as the minimal number (or fraction, in the normalized case) of additional edges needed to make a hypergraph a simplicial complex. 

Our formal definition uses the notion of an induced simplicial complex defined in Sec.~\ref{subsec:definitions}.
Given a hypergraph $\bH = (V, E)$ for which we want to measure the ES, we find its maximal edges $\widetilde{E}$ and construct the simplicial complex $\bS = (V, C)$ induced on $\bH$, with $C=\cup_{e \in \widetilde{E}} \ps{e}$. 
The edit simpliciality is then 
\begin{equation}
  \label{eq:ES}
  \ses = \frac{|E| - |\widetilde{E}|}{|C| - |\widetilde{E}|},
\end{equation}
again satisfying $\ses\in[0,1]$; see Fig.~\ref{fig:explanation_figure}C.
We subtract the number of maximal edges from the total number of edges so that the edit simpliciality is zero in the case of a hypergraph with no subfaces.
(We note that one can use the induced simplicial complex to define variants of the ES, e.g., a simplicial edit distance $\des = |C| - |E|$ or a normalized distance $\dnes = (|C| - |E|)/(|C| - |\widetilde{E}|) = 1 - (|E| - |\widetilde{E}|)/(|C| - |\widetilde{E}|) = 1 - \ses$.)

The ES answers a slightly different question than the SF does---it counts missing hyperedges that would make the dataset into a simplicial complex, rather than the edges that already satisfy downward closure.
It thus offers a complementary, equally interpretable measure of simpliciality.
However, the ES has the disadvantage of being sensitive to outliers, as a handful of large hyperedges with few inclusions will rapidly drive $\ses$ towards $0$.
Indeed, a hyperedge of size $m$ without any inclusion contributes one edge to $|E|$ but $2^m$ edges to $|C|$ in the denominator of Eq.~\eqref{eq:ES}. 

\vspace{0.1in}
\noindent \textbf{Face Edit Simpliciality}.
Finally, building upon the idea of edit simpliciality, we define a more localized notion of simpliciality, using the number of subfaces that must be added to the hypergraph to make a particular face a simplex.

Given a hyperedge $e$, the number of edges one must add to the hypergraph to make $e$ a simplex is 
\begin{equation*}
  d_{\mathrm{FES}}(e) = |\ps{e}| - |c|,  
\end{equation*}
where $c = \{ f \in E \mid f \subseteq e\}$.
We can think of this quantity as an edit distance, or \emph{face edit distance}.
We use this quantity to define an average 
\begin{equation*}
  \bar{d}_{\mathrm{FES}} = \frac{1}{\left| F\right|}\sum_{e\in F} d_{\mathrm{FES}}(e),
\end{equation*}
where $F$ is a set of edges---most commonly, $F = \widetilde{E}$ or $E$.
We exclusively use $F = \widetilde{E}$ in this study.
These quantities are on the scale of counts, and to define quantities analogous to previous simpliciality measures, we thus introduce a per-face normalization, either on a distance scale (meaning that the quantity grows as the dataset becomes less simplicial):
\begin{equation*}
  \bar{d}_{\mathrm{NFES}} = \frac{1}{\left| F\right|}\sum_{e\in F} \frac{d_{\mathrm{FES}}(e)}{\left|\ps{e}\right|-1},
\end{equation*}
or, similarly to previous definitions, on a simpliciality scale:
\begin{equation}
  \sfes = \frac{1}{\left| F\right|}\sum_{e\in F} \left(1 - \frac{d_{\mathrm{FES}}(e)}{\left|\ps{e}\right|-1}\right).
\end{equation}
We call this last measure the \emph{face edit simpliciality} (FES).
We subtract one in the denominators of both expressions so that when an edge has no subfaces, its normalized face edit distance is one.

The FES normalizes the face edit distance as a fraction of its maximal simpliciality. 
This normalization removes the dominance of large edges in the calculation of $\ses$ and, in fact, exponentially down-weights the contribution of these edges. 
In addition, because this metric is computed on faces, this is an averaged local metric.

% ==========================================
\subsection{Important considerations when measuring simpliciality}
\label{subsec:considerations} 
% ==========================================
Before we turn to applications in Sec.~\ref{sec:results}, let us discuss three design choices that may impact the conclusion we reach about the simpliciality of a dataset.

First, we note that the formal definition of a simplicial complex can be unnecessarily strict when used to represent perfect inclusion structures.
By definition, a simplex always contains singletons (edges comprising a single node) and the empty set. 
Several datasets will not include such interactions by construction.
One example is proximity datasets, where edges encode proximity events in which two or more nodes become in close contact during the observation period.
Because of their spatial nature, these datasets are often very dense and contain many inclusions~\cite{battiston_networks_2020}.
Yet, according to the standard definition, these will never be simplicial complexes due to the absence of singletons.
Another example is email datasets, which also do not contain singletons unless one includes emails that individuals send to themselves.
Because we define our notion of inclusion in terms of simplicial complexes, our measures will label these datasets as having no inclusion structure whatsoever.

To circumvent this issue, we use a relaxed definition of downward closure that excludes singletons wherever it makes sense.
The relaxation uses the notion of a \emph{size-restricted power set} $\srps{X}{K}$, where $K$ is a set of integers, defined as
\begin{equation}
  \label{eq:size_restricted_power_set}
  \srps{X}{K} = \left\{ x \in \ps{X} \ \Big| \ |x| \in K\right\}.
\end{equation}
For example, given an edge $e$ of size $n$, $\srps{e}{\{2,\dots,n-1\}}$ is the set of $2^{|e|} - |e| - 2$ subfaces of $e$ excluding the empty set, all singletons (sets of size one), and the edge $e$ itself.
Relaxed measures of simpliciality follow by substituting $\ps{X}$ for $\srps{X}{K}$ in all the measures of Sec.~\ref{subsec:measures}.
Hence, for example, we obtain a relaxation of $\ssf$ by replacing the definition of $S$, the set of the hyperedges of $\bH$ that are also simplices, by $S = \{e\subseteq E\mid\srps{e}{K}\subseteq E\} $), where $K=\big\{2,...,|e|\big\}$.

The results shown in Sec.~\ref{sec:results} are all calculated using size restrictions to exclude singletons and the empty set.
However, we note that this technique can be used more generally to exclude any interaction sizes deemed unimportant, anomalous, or problematic \cite{landry_filtering_2024};  or, conversely, to be more strict and to include singletons (say, when analyzing academic co-authorship networks, where single-author papers can meaningfully impact the inclusion structure of the dataset).

Second, we observe that special hyperedges we call ``minimal faces'' may significantly skew the simpliciality of a dataset.
The \emph{minimal faces} of a hypergraph $H$ are the edges of the minimal size, i.e., $|e| = \min(K)$, where $K$ is the set of sizes in the size-restricted powerset (In a traditional simplicial complex, the minimal faces are singletons).
With the size restrictions in place, the minimal faces of a hypergraph are always simplices because, by definition, there are no smaller edges for these edges to include.
We argue that when measuring the simpliciality of a dataset, it is most meaningful to focus on the faces for which inclusion is \emph{possible}, and so we exclude these minimal faces when counting potential simplices.

Note that this design choice operates differently from the size restriction imposed by the modified power set introduced in Eq.~\eqref{eq:size_restricted_power_set}; in that context, we argued for ignoring edges that can prevent other edges from being simplices, while here we suggest that counting minimal faces as potential simplices will strongly affect the value of simpliciality.
Our strategy is as follows.
For SF, this means that both $S$ and $E$ exclude the minimal-sized edges. For ES,  we exclude maximal faces that are also minimal faces when constructing the minimal simplicial complex. 
And for FES, we only average over maximal edges that are not minimal faces.

Third and finally, since the number of potential subfaces of a hyperedge grows exponentially with its size, computational issues prevent us from applying our measures to large hyperedges.
For this reason, we select a maximum face size $k$ (we use $k=11$ throughout), again using the size restriction to define our metrics.
This drops information about large hyperedges but speeds up computation drastically in practical applications.

% ==========================================
\subsection{Local simpliciality}
\label{subsec:local-simpliciality}
% ==========================================

Simpliciality, up to this point a global metric, can also be localized on a smaller subset of the higher-order network to yield information about its local structure.
The various face-centric measures used in our construction of the FES provide this information at the level of faces.
But for more flexibility, we also use subhypergraphs to define \emph{nodal simpliciality} measures of our global measures.
More specifically, given a hypergraph $\bH = (V,E)$ and a node $v\in V$, we define the neighborhood of $v$ as $n(v) = \{u \in V \mid u, v \in e\in E\}$ and the associated subsets $\widehat{V}=v\,\cup\, n(v)$ and $\widehat{E} = \{e \in E \mid e\subseteq \widehat{V}\}$. Then the simpliciality of node $v$ is the simpliciality defined on the subhypergraph $\widehat{\bH} =(\widehat{V}, \widehat{E})$ induced on the neighborhood of $v$. Note that when $v$ is an isolated node or when $\widehat{E}$ only contains minimal faces and we do not consider these potential simplices, the nodal simpliciality will be undefined.

% ==========================================
\section{Results} 
\label{sec:results} 
% ==========================================

\begin{table*}[t]
\centering
\begin{tabular}{lccccccccc}
Dataset&\null\quad\null & $|V|$ & $|E|$ & $\langle k\rangle$ & $\langle s\rangle$ &\null\quad\null & $\ssf$ & $\ses$ & $\sfes$\\
\hline\\
\textbf{Proximity datasets} &&&&&&&&&\\
\texttt{contact-primary-school} & & 242 & 12,704 & 52.50 & 2.42 & & 0.85 & 0.88 & 0.94 \\
\texttt{contact-high-school} & & 327 & 7,818 & 23.91 & 2.33 & & 0.81 & 0.91 & 0.92 \\
\texttt{hospital-lyon} & & 75 & 1,824 & 24.32 & 2.43 & & 0.91 & 0.94 & 0.97 \\[0.1in]
\textbf{Email datasets} &&&&&&&&&\\
\texttt{email-enron} & & 143 & 1,442 & 10.08 & 2.97 & & 0.31 & 0.04 & 0.50 \\
\texttt{email-eu} & & 967 & 23,729 & 24.54 & 3.12 & & 0.32 & 0.04 & 0.52 \\[0.1in]
\textbf{Biological datasets} &&&&&&&&&\\
\texttt{diseasome} & & 516 & 314 & 0.61 & 3.00 & & 0.00 & 0.02 & 0.04 \\
\texttt{disgenenet} & & 1,982 & 760 & 0.38 & 5.14 & & 0.00 & 0.00 & 0.01 \\
\texttt{ndc-substances} & & 2,740 & 4,754 & 1.74 & 5.16 & & 0.02 & 0.00 & 0.07 \\[0.1in]
\textbf{Other} &&&&&&&&&\\
\texttt{congress-bills} & & 1,715 & 58,788 & 34.28 & 4.95 & & 0.03 & 0.00 & 0.10 \\
\texttt{tags-ask-ubuntu} & & 3,021 & 145,053 & 48.01 & 3.43 & & 0.15 & 0.11 & 0.46 \\[0.1in]
\hline\\
\end{tabular}
\caption{Properties of empirical datasets and their simpliciality. $|V|$, $|E|$, $\langle k\rangle$, $\langle s\rangle$, $\ssf$, $\ses$, and $\sfes$ denote the number of nodes, the number of hyperedges, the mean degree, the mean edge size, the simplicial fraction (SF), edit simpliciality (ES), and the face edit simpliciality (FES), respectively.}
\label{tab:network_stats}
\end{table*}

% ==========================================
\subsection{Empirical datasets}
\label{subsec:data} 
% ==========================================

As the first demonstration of the simpliciality measures, we analyze empirical higher-order datasets from several general domains. 
All datasets are obtained from the \texttt{xgi-data} repository~\cite{landry_xgi-data_2023} and are openly available.
Following the considerations highlighted in Sec.~\ref{subsec:considerations}, we preprocess these datasets to remove singletons, multiedges, and isolated nodes. 
In addition, for computational feasibility, we only consider hyperedges of size 11 (order, defined as the size minus one, of 10) and smaller.
Basic structural properties of the pre-processed datasets are shown in Table~\ref{tab:network_stats}.

\begin{figure*}[t]
    \centering
    \includegraphics[width=0.8\linewidth]{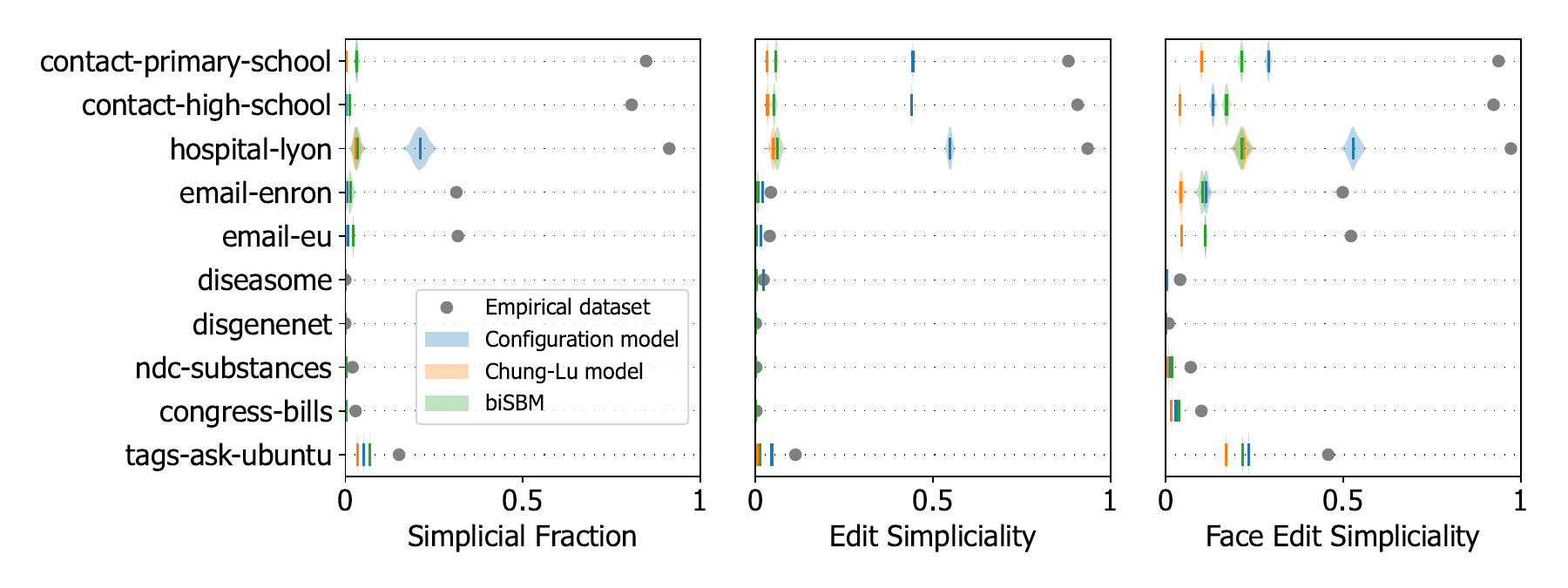}
    % scaling is important to keep font size consistent
    \caption{\csentence{The simpliciality of empirical datasets and their fitted generative models.} The simpliciality of empirical datasets compared with samples from three higher-order generative models: the hypergraph configuration model \cite{chodrow_configuration_2020}, the bipartite Chung-Lu model \cite{aksoy_measuring_2017}, and the bipartite degree-corrected stochastic block model \cite{yen_community_2020}. The violin plots indicate the simpliciality of samples drawn from the fitted generative models, and the solid vertical lines indicate the mean of the distributions. Empirical results are shown with a solid circle.}
    \label{fig:data_vs_generative_models}
\end{figure*}

Our sample of datasets contains various types of complex systems. 
We analyze three proximity datasets~\cite{benson_simplicial_2018,mastrandrea_contact_2015,stehle_high-resolution_2011,vanhems_estimating_2013,landry_xgi-data_2023} (\texttt{contact-primary-school}, \texttt{contact-high-school}, and \texttt{hospital-lyon}), which are collected via proximity sensors with a range of roughly 1 meter. Nodes are individuals, and an edge is created from a proximity event, where two individuals are closer than 1 meter apart. To create a higher-order dataset, each maximal clique is converted into a hyperedge at each time step. Unique to proximity datasets are their geometrical constraints, and because of the proximity sensor range, 5-hyperedges are the largest edges present in these datasets.
We also include two datasets of email interactions~\cite{benson_simplicial_2018,klimt_enron_2004,yin_local_2017,leskovec_graph_2007}: \texttt{email-enron} and \texttt{email-eu}. In both cases, the nodes are email addresses, and the hyperedges are emails, at the defunct company Enron in the former case and a large European research institution in the latter.
Three datasets are loosely associated with biological 
processes~\cite{benson_simplicial_2018,goh_human_2007,pinero_disgenet_2020}: \texttt{diseasome}, \texttt{disgenenet}, and \texttt{ndc-substances}. In these datasets, nodes are compounds, diseases, or genes, while hyperedges are interactions amongst these to represent pharmaceuticals, symptoms, and diseases. 
Finally, we include two miscellaneous datasets as well: \texttt{tags-ask-ubuntu}~\cite{benson_simplicial_2018} higher-order dataset in which a node is a tag on Stack Overflow, and an edge is a question to which the tags are associated; and the \texttt{congress-bills} dataset~\cite{benson_simplicial_2018,fowler_connecting_2006,fowler_legislative_2006} where nodes are congresspeople and edges represent the sponsoring and co-sponsoring congresspeople for a particular bill.

Numerical values of the simpliciality measures are shown in Table~\ref{tab:network_stats} for all of these datasets. We find that values for simpliciality fill the spectrum from 0 to 1, depending on the data. The proximity datasets have large simpliciality for all three measures, while the biological datasets have low simpliciality for all three measures. The email datasets have a very small ES simpliciality, with moderate simpliciality for the other two measures. (And since we use size restrictions to exclude singletons, the lack or absence of emails sent to oneself does not affect this assessment.) Similarly, the \texttt{tags-ask-ubuntu} dataset has a range of simpliciality values depending on which measure we consider. This shows that the measures we have defined in section~\ref{subsec:measures} capture different features of the inclusion structure.

While the measures give differing perspectives on the simpliciality of each dataset, we verify that they broadly agree with a correlation analysis.
The Pearson correlation coefficient is $\rho=0.97$ between the SF and ES, $\rho=0.95$ between the SF and FES, and $\rho=0.90$ between the ES and FES (all significant at the $p=0.001$  level). Hence, the values are closely and linearly related in our sample.
They also order datasets similarly, from the least to most simplicial, since the Spearman rank-order correlation coefficient is $\rho=0.89$ between SF and ES, $\rho=0.997$ between SF and FES, and $\rho=0.90$ between ES and FES (all significant at the same level).

Although our correlation analysis confirms that these measures roughly capture the same concepts, the datasets where they depart from one another highlight their key differences.
In our experiments, these differences are due to features such as large edges with few included edges, many edges that are mostly closed downward, and different edge size distributions.
Networks of organizational email messages are examples of the first case, where very large organization- or department-wide emails may be sent with no guarantee that emails are also sent between every possible subgroup of individuals.
In this case, we would expect ES to be extremely low while the SF need not be low.
Proximity networks are examples of the second case, i.e., dense downward-closed datasets.
We see this by noting that the SF is not 1, while both ES and FES are close to 1 due to the SF penalizing almost-simplicial edges.
Lastly, the edge size distribution has strong implications on all measures; for the same average edge size and number of edges, increasing both the number of small and large edges will affect the SF and ES measures.
For ES, the large edges will exponentially increase the number of subfaces needed to create a simplicial complex, driving the simpliciality to zero.
In contrast, increasing the number of small edges can create more small simplices, increasing SF.

% ==========================================
\subsection{Generative models of higher-order networks}
\label{subsec:generative-models} 
% ==========================================

To complement our analysis of empirical data, we also examine the simpliciality of synthetic data generated with generative models for higher-order networks.

We focus on models of hypergraphs designed to describe and analyze arbitrary higher-order structures.
There are several random hypergraph models, including, among many classes of models, preferential attachment models \cite{wang_evolving_2010,avin_random_2019,do_structural_2020,barthelemy_class_2022}, models with community structure \cite{aksoy_measuring_2017,zhang_exact_2023,kim_stochastic_2018,chodrow_generative_2021,ruggeri_community_2023}, models with specified degree and size sequences \cite{chodrow_configuration_2020,aksoy_measuring_2017}, \ER models \cite{dewar_subhypergraphs_2018,iacopini_simplicial_2019}, models with latent nodal variables governing edge formation \cite{di_gaetano_percolation_2024}, and geometric models \cite{barthelemy_class_2022,turnbull_latent_2023,lunagomez_geometric_2017}.
Higher-order random models that are commonly fit to empirical datasets include the configuration model \cite{chodrow_configuration_2020}, the bipartite Chung-Lu model \cite{aksoy_measuring_2017,chodrow_configuration_2020}, and the bipartite degree-corrected stochastic block model \cite{yen_community_2020}. See Ref.~\cite{battiston_networks_2020} for an extensive overview.
Overwhelmingly, generative hypergraph models lack explicit control over the inclusion structure of hyperedges, so there are often relatively few simplices.

We focus our analysis on three models: the configuration model \cite{chodrow_configuration_2020}, the bipartite Chung-Lu model \cite{aksoy_measuring_2017}, and the bipartite degree-corrected stochastic block model (biSBM) \cite{yen_community_2020}.

We fit each model to the empirical datasets of Table~\ref{tab:network_stats}, use the fitted models to generate a distribution of higher-order networks (the \emph{posterior predictive} distribution in Bayesian parlance), and analyze the resulting distribution of simpliciality values.

In all cases,  when sampling synthetic higher-order networks from the three generative models, we generate $10^3$ realizations of each model for each empirical dataset. 
We use the double edge-swap algorithm presented in Ref.~\cite{chodrow_configuration_2020} to sample from the configuration model and performed $10 \times |E|$ edge swaps, roughly in accordance with \cite{dutta_sampling_2023}.
For the bipartite Chung-Lu model \cite{aksoy_measuring_2017}, we extract the degree and edge size sequences and then use a bipartite variation of the algorithm introduced in Ref.~\cite{miller_efficient_2011} and available in XGI \cite{landry_xgi_2023} to sample from this model.
Lastly, when sampling from the biSBM, we used a Markov chain Monte Carlo method with a bipartite prior using the algorithm described in Ref.~\cite{yen_community_2020}.

All results are reported in Fig.~\ref{fig:data_vs_generative_models}.
Overwhelmingly, we see that the generative models cannot accurately capture the simpliciality of datasets when they have a non-trivial inclusion structure.
While it does not reproduce the correct values, the hypergraph configuration model consistently captures the inclusion structure better than the biSBM and the bipartite Chung-Lu model, irrespective of the simpliciality measure used.
This may be due to the exact specification of the degree and edge size sequences; the Chung-Lu model and biSBM only match these sequences in expectation.

% ==========================================
\subsection{Local measures of simpliciality}
\label{subsec:local} 
% ==========================================

\begin{figure}
    \centering
    \includegraphics[width=0.9\linewidth]{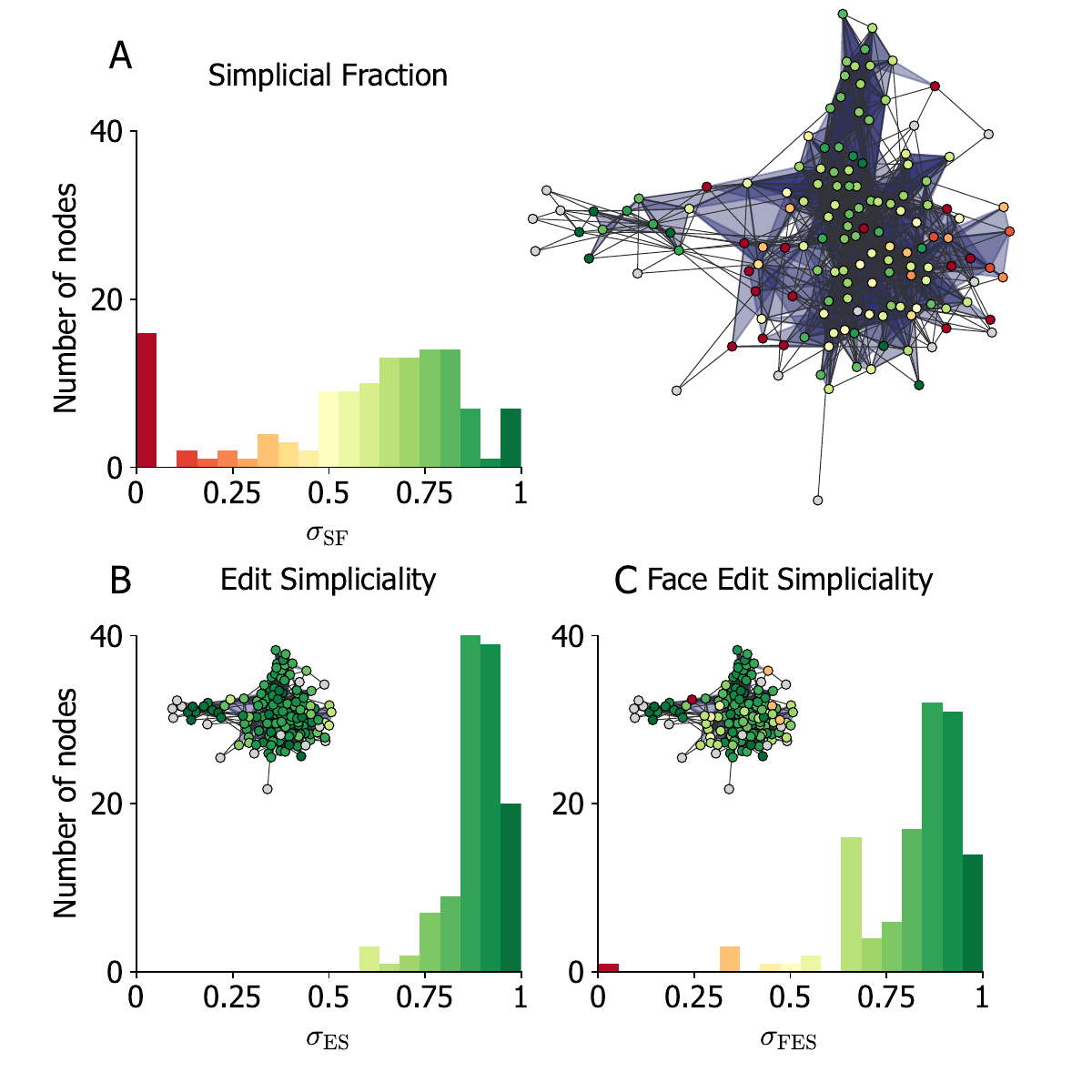}
    \caption{\csentence{The local simpliciality of an empirical dataset.} The local (A) simplicial fraction, (B) edit simpliciality, and (C) face edit simpliciality of the \texttt{email-enron} dataset filtered to include interactions of sizes 2 and 3. The colors of the histogram bars match the node colors on their corresponding network visualization. Nodes for which the local simpliciality is undefined are colored in grey.}
    \label{fig:email-enron_local_simpliciality}
\end{figure}
As a final demonstration, we apply our local measures of simpliciality to the dataset of emails sent by Enron employees (142 nodes and 1126 hyperedges, filtered to include interactions of sizes 2 and 3).  Results are shown in Fig.~\ref{fig:email-enron_local_simpliciality}.

Focusing on the histograms first, we find that the SF has the most variability and that the FES covers a similarly large range. In contrast, the ES tells us that nearly every neighborhood is strongly simplicial.
This is expected behavior because the ES relies on a simplicial complex induced on the ego-hypergraph; the size of the largest hyperedges in this ego-hypergraph can be substantially smaller than that of the largest hyperedges in the whole hypergraph.
As a result, the denominator of Eq.~\eqref{eq:ES} is reduced, increasing the local ES systematically.
In contrast, when we take a subset of nodes to form an ego-hypergraph, it is easy to omit a small subface shared by many hyperedges, thus leading to a very small SF (and, similarly, to a small FES).

Turning to the spatial distribution of simpliciality shown in the insets, we see that the SF and FES find a region of high simpliciality at the network's core with regions of low simpliciality on its edges.
In fact, these two measures are largely in agreement, with a Pearson correlation coefficient of $\rho=0.84$ between the local SF and FES. (The correlation drops to $\rho=0.6$ when comparing the SF and ES).
We also observe several nodes for which simpliciality is undefined.
These nodes are only connected via minimal faces in their ego-hypergraphs, and these faces are excluded when calculating both potential and actual simplices.

Finally, inspecting Fig.~\ref{fig:email-enron_local_simpliciality}, we notice that, in this case, nodes of similar simpliciality tend to be connected to one another.
To quantify this observation, we define the \textit{simplicial assortativity} as the Pearson correlation coefficient of the simpliciality of pairs of nodes connected by at least one hyperedge. 
More formally, we use the unweighted adjacency matrix of the hypergraph, $\mathbf{A}$, defined as
\begin{equation*}
A_{ij} = \begin{cases} 1, & \bigl[\mathbf{BB}^{\mathrm{T}}\bigr]_{ij} > 0\text{ and } i\neq j,\\ 0,& \text{ otherwise,} \end{cases}
\end{equation*}
where $B$ is the incidence matrix of the hypergraph, such that $B_{ij}=1$ if node edge $j$ is incident on node $i$. 
The simplicial assortativity, $\rho$, can then be defined as
\begin{equation}
\rho = \sum_{i,j}\frac{A_{ij}(\sigma_i -E[\sigma])(\sigma_i -E[\sigma])}{\text{Var}[\sigma]},
\end{equation}
where $\sigma_i$ is the local simpliciality of node $i$ according to one of our measures. The simplicial assortativity for SF, ES, and FES are denoted $\rho_{\mathrm{SF}}$, $\rho_{\mathrm{ES}}$, and $\rho_{\mathrm{FES}}$ respectively.
This coefficient is equivalent to the assortativity coefficient~\cite{newman_mixing_2003} of the local simpliciality on the unweighted pairwise projection of the hypergraph.
\begin{table}
\begin{center}
\begin{tabular}{lcccc}
Dataset &\null\quad\null  & $\rho_{\mathrm{SF}}$ & $\rho_{\mathrm{ES}}$ & $\rho_{\mathrm{FES}}$ \\ 
\hline\\
\textbf{Proximity datasets} &&&&\\
\texttt{contact-primary-school} && 0.15 & 0.17 & 0.14 \\ 
\texttt{contact-high-school} && 0.22 & 0.37 & 0.24 \\ 
\texttt{hospital-lyon} && -0.02 & -0.01 & -0.01 \\[0.1in]
\textbf{Email datasets} &&&&\\
\texttt{email-enron} && 0.29 & 0.29 & 0.24 \\ 
\texttt{email-eu} && 0.19 & 0.16 & 0.16 \\[0.1in]
\textbf{Biological datasets} &&&&\\
\texttt{ndc-substances} && 0.56 & 0.54 & 0.69 \\ 
\texttt{diseasome} && N/A & 0.28 & 0.68 \\ 
\texttt{disgenenet} && N/A & 0.28 & 0.78 \\[0.1in]
\textbf{Other} &&&&\\
\texttt{congress-bills} && 0.78 & 0.33 & 0.75 \\ 
\texttt{tags-ask-ubuntu} && -0.03 & -0.24 & 0.04 \\[0.1in]
\hline\\
\end{tabular}
\caption{\label{tab:simplicial_assortativity} The simplicial assortativity of each dataset filtered to only include interactions of sizes two and three for computational tractability.}
\end{center}
\end{table}

One should expect local simpliciality to be assortative as any given subface contributes to the simpliciality of all their nodes.
Table~\ref{tab:simplicial_assortativity} shows that the situation is a bit more nuanced.
For \texttt{tags-ask-ubuntu}, FES is weakly assortative, SF is weakly disassortative, and ES is strongly disassortative. 
It is particularly striking that despite the \texttt{hospital-lyon} dataset being highly simplicial (as seen in Fig.~\ref{fig:data_vs_generative_models}), it is also weakly disassortative.

% ==========================================
\section{Conclusion}
\label{sec:discussion}
% ==========================================

In this paper, we have introduced measures to summarize the inclusion structure---the simpliciality---of hypergraphs.
We have presented three measures of simpliciality but recognize that other definitions of simpliciality may also prove useful.
We have discussed how the simpliciality of higher-order datasets depends on many factors, including, but not limited to, the manner in which the dataset was collected, its domain, and the measure of simpliciality.
When fitting common generative models to several empirical higher-order networks, we found that the simpliciality of the original dataset is often much higher than the simpliciality of the posterior predictive distribution of fitted models by any measure of simpliciality.
Measuring the simplicial assortativity indicates that the simpliciality displays different levels of localization.

We hope this study will serve as a starting point for network scientists aiming to characterize higher-order network datasets and look forward to future work developing these methods along a number of dimensions of interest. 

First,  we presented global- and node-level definitions of simpliciality, but other scales of interaction may yield further insights into the inclusion structure of the data~\cite{mariani_nestedness_2019}.
Future work could explore mesoscale measures of simpliciality that describe how, for example, simpliciality varies between communities.
One could also obtain the largest simplicial component or the set of simplicial components in a hypergraph.
In addition, we have restricted ourselves to unweighted simplicial complexes, but one might consider extending these notions to weighted simplicial complexes \cite{baccini_weighted_2022}. 

Second, our approach complements the existing literature on nestedness in bipartite networks~\cite{mariani_nestedness_2019}, which shows that nestedness exists for a wide variety of unipartite and bipartite networks~\cite{johnson_factors_2013}.
Existing work shows that nestedness is important for the function of networks in many domains~\cite{bastolla_architecture_2009,saavedra_strong_2011,kamilar_cultural_2014,cantor_nestedness_2017}, and comparing these findings from the perspective of simpliciality could offer additional insights from both a structural and mechanistic perspective.

Finally, we have shown a disparity between the simpliciality of artificial datasets and observed ones.
Our findings should thus inform new higher-order network models that specify the inclusion structure of the network and can be fit to empirical higher-order datasets.

\section*{Abbreviations}
\begin{itemize}
    \item \textbf{SF}: Simplicial fraction
    \item \textbf{ES}: Edit simpliciality
    \item \textbf{FES}: Face edit simpliciality
    \item \textbf{biSBM}: bipartite degree-corrected stochastic block model
\end{itemize}

\newpage
\appendix

\section{\label{sec:appendix-cm-convergence} Convergence of the configuration model MCMC}

\begin{figure}
    \centering
    \includegraphics[width=\textwidth]{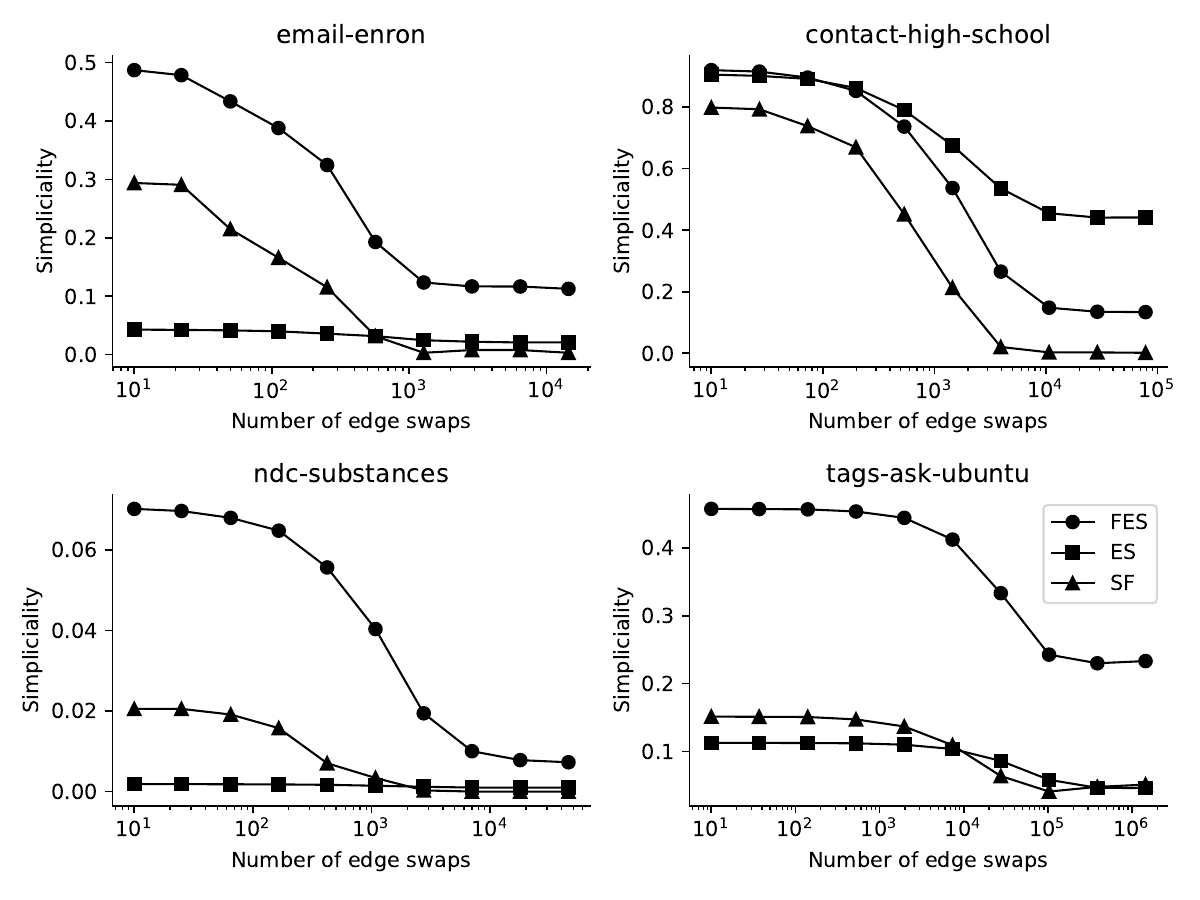}
    \caption{Convergence of the simpliciality for representative datasets from each data domain. The rightmost markers in all plots correspond with the number of edge swaps used when generating the results in Fig.~\ref{fig:data_vs_generative_models}.}
    \label{fig:cm_convergence}
\end{figure}

To sample from the configuration model, we implemented the Markov chain Monte Carlo algorithm proposed in Ref.~\cite{chodrow_configuration_2020}.
At each step, (1) two edges are selected at random, $e_1$ and $e_2$; (2) two nodes are selected at random, $i \in e_1$ and $j \in e_2$; and (3) the memberships of these nodes are swapped to form two new edges, $\widetilde{e}_1 = (e_1 \setminus i) \cup j$ and $\widetilde{e}_2 = (e_2 \setminus j) \cup i$. This operation is accepted if the move does not create loopy hyperedges, i.e., $|e_1| = |\widetilde{e}_1|$ and  $|e_2| = |\widetilde{e}_2|$. In Ref.~\cite{chodrow_configuration_2020}, the criterion for convergence is left for future work, and Ref.~\cite{dutta_sampling_2023} presents criteria for the convergence of the configuration model for pairwise networks. In Figure~\ref{fig:cm_convergence}, we show that the number of double-edge swaps chosen for our configuration model algorithm ($10\times|E|$) is sufficient for convergence with respect to all measures of simpliciality.

From Fig.~\ref{fig:cm_convergence}, we see that the configuration model sampler has roughly achieved the stationary value of simpliciality after roughly 10\% of the edge swaps have been completed. Assessing convergence in a statistically robust manner is necessary to ensure uniform sampling from the hypergraph configuration model, but this heuristic is sufficient for our purposes.

\section{\label{sec:appendix-algorithms} Measuring simpliciality efficiently}

For all measures, we leverage the trie data structure \cite{fredkin_trie_1960} to efficiently compute the measures described in this paper. The trie structure allows us to efficiently verify whether a subface exists ($\mathcal{O}(|e|)$, for an edge $e$). To be compatible with the trie data structure, when adding an edge to the trie and when searching for an edge in the trie, we first sort the nodes in that edge. Below, we present the algorithms employed in generating all results.

\paragraph{Simplicial fraction.} For each edge, if a single subface is absent, we can immediately determine that the edge is not a simplex. This can be very efficient for sparse hypergraphs.

\paragraph{Edit simpliciality.} There are two ways to compute the edit simpliciality: the exhaustive method and a more memory-efficient version. The exhaustive method is simpler and more computationally efficient. However, the memory requirements are enormous because it stores every missing subface. The memory-efficient version keeps track of the number of missing subfaces, not the missing subfaces themselves. This leverages the fact that the number of hyperedges intersecting with any given hyperedge is typically much smaller than the number of hyperedges. We store the hypergraph of maximal edges for fast retrieval of the neighbors of a given maximal face, which is needed in Algorithm~\ref{alg:es_mem_efficient}.

\paragraph{Face edit simpliciality.} This computation is more straightforward than the ES computation because it is a local measure that requires relatively little memory and allows us to compute the number of missing subfaces and then store the FES as a running average.

\newpage

\begin{algorithm}[H]
    \caption{Simplicial Fraction}
    \label{alg:sf}
    \DontPrintSemicolon
    \KwIn{$K$, a set of edge sizes}
    \custominput{$m$, the minimum acceptable simplex size}
    \custominput{$\bH = (V, E)$, a hypergraph}
    \custominput{$T$, a trie constructed from the edges in $\bH$}
    \KwOut{$\ssf$}
    $\ssf = 0$\;
    $F = \{e \in E \mid |e| \in K, |e| \geq m\}$\;
    \tcp{Iterate over eligible simplices.}
    \For{$f \in F$}
    {
        IsSimplex = true\;
        \For{$e \in \srps{f}{K}$}
        {
            \tcp{If a single subface is absent, the edge is not a simplex.}
            \If{$e \notin T$}
            {
                IsSimplex = false\;
                \Break\;
            }
        }
        \tcp{Update the fraction of simplices.}
        \If{\textup{IsSimplex}}
        {
            $\ssf \leftarrow \ssf + 1/|F|$\;
        }
    }
    \KwRet $\ssf$\;
\end{algorithm}

\begin{algorithm}
    \caption{Exhaustive Edit Simpliciality}
    \label{alg:es_exhaustive}
    \DontPrintSemicolon
    \KwIn{$K$, a set of edge sizes}
    \custominput{$m$, the minimum acceptable simplex size}
    \custominput{$\bH = (V, E)$, a hypergraph}
    \custominput{$T$, a trie constructed from the edges in $\bH$}
    \KwOut{$\ses$}
    $\ses = 0$\;
    $E = \{e \in E \mid |e| \in K, |e| \geq m\}$\;
    \tcp{Construct the set of maximal faces.}
    $F = \{e \in E \mid e \notin f, \ \forall f\in E\}$\;
    \tcp{D stores the unique missing subfaces.}
    $D = \emptyset$, is a set of sets\;
    \tcp{Iterate over all maximal faces.}
    \For{$f \in M$}
    {
        \tcp{For each maximal face of the hypergraph, add all of its missing subfaces not already present in the global set of missing faces.}
        \For{$e \in \srps{f}{K}$}
        {
            \If{$e \notin T$}
            {
                $D \leftarrow D \cup e$\;
            }
        }
    }
    \tcp{The number of edges in the minimal simplicial complex is the sum of the number of edges in the original hypergraph and the number of missing subfaces.}
    $\ses = (|E| - |\widetilde{E}|)/(|E| - |\widetilde{E}| + |D|)$\;
    \KwRet $\ses$\;
\end{algorithm}

\begin{algorithm}
    \caption{Memory-Efficient Edit Simpliciality}
    \label{alg:es_mem_efficient}
    \DontPrintSemicolon
    \KwIn{$K$, a set of edge sizes}
    \custominput{$m$, the minimum acceptable simplex size}
    \custominput{$\bH = (V, E)$, a hypergraph}
    \custominput{$T$, a trie constructed from the edges in $\bH$}
    \KwOut{$\ses$}
    $\ses = 0$\;
    $E = \{e \in E \mid |e| \in K, |e| \geq m\}$\;
    \tcp{Construct the set of maximal faces.}
    $F = \{e \in E \mid e \notin f, \ \forall f\in E\}$\;
    $d = 0$\;
    \tcp{Iterate over all enumerated maximal faces.}
    \For{$i=1\dots |F|$}
    {
        $f = F_i$\;

        \tcp{First, calculate the number of missing faces for a given maximal face.}
        $\widetilde{d} = |\srps{f}{K}|$\;
        \For{$e \in \srps{f}{K}$}
        {
            \If{$e \in T$}
            {
                $\widetilde{d} \leftarrow \widetilde{d} - 1$\;
            }
        }
        \tcp{Update the total number of missing subfaces}
        $d \leftarrow d + \widetilde{d}$\;
        \tcp{Calculate the number of redundant missing subfaces counted for the maximal faces already seen. To prevent looping over all previous maximal edges, we iterate only over the previous maximal faces, which are also neighbors of the current maximal face.}
        $D = \emptyset$\;
        \For{$j = \{1\dots i - 1\}\cap \{k \mid e_k \cap f \neq \emptyset\}$}
        {
            \tcp{For each prior maximal face, we add the missing edges formed by the powerset of the intersection of that edge and the current maximal edge to the complete set of redundant missing edges.}
            $e = F_j$\;
            $g = e\cap f$\;
            \For{$h \in \srps{g}{K\cup |g|}$}
            {
                \If{$g \notin T$}
                {
                    $D \leftarrow D \cup g$\;
                }
            }
        }
        \tcp{Subtract the redundant missing subfaces}
        $d \leftarrow d - |D|$
    }
    \tcp{The number of edges in the minimal simplicial complex is the sum of the number of edges in the original hypergraph and the number of missing subfaces.}
    $\ses = (|E| - |\widetilde{E}|)/(|E| - |\widetilde{E}| + d)$\;
    \KwRet $\ses$\;
\end{algorithm}

\begin{algorithm}[H]
    \caption{Face Edit Simpliciality}
    \label{alg:fes}
    \DontPrintSemicolon
    \KwIn{$K$, a set of edge sizes}
    \custominput{$m$, the minimum acceptable simplex size}
    \custominput{$\bH = (V, E)$, a hypergraph}
    \custominput{$T$, a trie constructed from the edges in $\bH$}
    \KwOut{$\sfes$}
    $\sfes = 0$\;
    $E = \{e \in E \mid |e| \in K, |e| \geq m\}$\;
    \tcp{Construct the set of maximal faces.}
    $F = \{e \in E \mid e \notin f, \ \forall f\in E\}$\;
    $\sfes = 0$\;
    \tcp{Iterate over all maximal faces.}
    \For{$f \in F$}
    {
        \tcp{For each maximal face, calculate the fraction of missing faces.}
        $s = 0$\;
        \For{$e \in \srps{f}{K}$}
        {
            \If{$e \in T$}
            {
                $s\leftarrow s + 1/(|\srps{f}{K}|-1)$\;
            }
        }
        \tcp{Update the running average.}
        $\sfes \leftarrow \sfes + s/|F|$\;
    }
    \KwRet $\sfes$\;
\end{algorithm}

%%%%%%%%%%%%%%%%%%%%%%%%%%%%%%%%%%%%%%%%%%%%%%
%%                                          %%
%% Backmatter begins here                   %%
%%                                          %%
%%%%%%%%%%%%%%%%%%%%%%%%%%%%%%%%%%%%%%%%%%%%%%
\newpage
\begin{backmatter}
\section*{Availability of data}
All datasets are available in the \texttt{xgi-data} repository~\cite{landry_xgi-data_2023}. The code used to generate all results and figures utilizes the XGI library~\cite{landry_xgi_2023} and is openly available on \href{https://github.com/nwlandry/the-simpliciality-of-higher-order-networks}{Github}~\cite{landry_code_2025}.

\section*{Competing interests}
The authors declare that they have no competing interests.

\section*{Funding}
N.W.L. and J.-G.Y. acknowledge support from the National Institutes of Health 1P20 GM125498-01 Centers of Biomedical Research Excellence Award. N.E. acknowledges the Harris Faculty Fellowship from Grinnell College.

\section*{Author's contributions}
N.W.L. conceived the project; N.W.L., J.G.Y., and N.E. designed the research; N.W.L. and N.E. performed the research; N.W.L., J.G.Y., and N.E. wrote the article; and N.W.L., J.G.Y., and N.E. edited the draft.

\section*{Acknowledgements}

N.W.L. would like to acknowledge the participants of the "Workshop on Modelling and Mining Complex Networks as Hypergraphs" at Toronto Metropolitan University and Tim LaRock for helpful conversations. N.W.L. would also like to thank Tzu-Chi Yen for lending his expertise on the biSBM inference. We thank François Th\'{e}berge for pointing out inconsistencies between the code and notation in the earlier versions of this manuscript.

% if your bibliography is in bibtex format, use those commands:
\bibliographystyle{bmc-mathphys} % Style BST file (bmc-mathphys, vancouver, spbasic).
\bibliography{references}

\end{backmatter}
\end{document}